\documentclass[prb,aps,twocolumn,amsmath]{revtex4}

\usepackage{graphicx}
\usepackage{epsfig}
\begin{document}
\title{Efficient electron spin detection with positively charged quantum dots}
\author{K. G\"{u}ndo\u{g}du, K. C. Hall, Thomas F. Boggess}
\affiliation{Department of Physics and Astronomy and Optical Science and Technology Center, The University of Iowa, Iowa City,\\
Iowa 52242}
\author{D. G. Deppe, O. B. Shchekin}
\affiliation{Microelectronics Research Center, Department of
Electrical and Computer Engineering, The University of Texas at
Austin, Austin, Texas 78712-1084}

\begin{abstract}
We report the application of time- and polarization-resolved
photoluminescence up-conversion spectroscopy to the study of spin
capture and energy relaxation in positively and negatively
charged, as well as neutral, InAs self-assembled quantum dots.
When compared to the neutral dots, we find that carrier capture
and relaxation to the ground state is much faster in the highly
charged dots, suggesting that electron-hole scattering dominates
this process. The long spin lifetime, short capture time, and high
radiative efficiency of the positively charged dots, indicates
that these structures are superior to both quantum well and
neutral quantum dot light-emitting diode (LED) spin detectors for
spintronics applications.
\end{abstract}

\maketitle

The efficient detection of spin-polarized carriers is a crucial
issue for the design of semiconductor-based spintronic
devices.\cite{Wolf:2001}  A light-emitting diode (LED)
configuration employing a quantum well as an optical marker has
proven to be an effective means of detecting spin polarized
carriers.\cite{Fiederling:1999,Ohno:1999}
 In a spin LED, the degree of circular polarization of the quantum
well luminescence provides a direct measure of the spin
polarization of the carriers arriving at the spatial location of
the quantum well.

The application of semiconductor quantum dots (QDs) in such a spin
detection scheme is expected to provide a substantial improvement
in spin sensitivity over the use of a quantum
well.\cite{Paillard:2001,Gotoh:1998,Chye:2002,Pryor:2003} Recent
studies of electron spin dynamics in neutral
QDs\cite{Paillard:2001,Gotoh:1998,Kalevich:2001} have revealed
that the discrete energy levels in quantum dots arising from
three-dimensional quantum confinement blocks the dominant spin
relaxation channels present in higher dimensional bulk and quantum
well systems, resulting in considerably longer spin relaxation
times. Combined with the high optical luminescence efficiency
observed in QDs,\cite{Arakawa:1986,Mao:1997,Shchekin:2000} these
long spin relaxation times should lead to larger spin-dependent
luminescence signatures in spin detection applications
incorporating QDs as an optical marker. The first spin LED using
neutral QDs was recently demonstrated.\cite{Chye:2002}

Few experiments have examined electron spin dynamics in charged
quantum dots.\cite{Cortez:2002,Kozin:2002} Through a comparison of
spin capture and relaxation dynamics in neutral, positively (+QDs)
and negatively (-QDs) charged QDs, we demonstrate that +QDs act as
a highly efficient detector for spin-polarized electrons. Our room
temperature time-resolved measurements reveal that, following
capture of spin-polarized electrons, the initial degree of
circular polarization of the QD ground state luminescence is more
than four times larger in +QDs compared to neutral QDs.  The
larger spin signature in +QDs originates from an increased rate of
capture of spin polarized electrons through interaction with the
built-in hole population,\cite{CommentNextQDPaper,Sosnowski:1998}
as indicated by the early time dynamics of the QD
photoluminescence. Rapid electron capture into +QDs reduces spin
relaxation in the GaAs barriers prior to capture, resulting in a
six-fold enhancement in the time-integrated spin detection
efficiency with the incorporation of positive charge on the QDs.
Our experiments, which represent the first measurement of electron
spin dynamics in positively-charged quantum dot nanostructures,
also indicate that the presence of a large population of excess
holes has little if any influence on the electron spin relaxation
time in the QDs, in contrast with the findings in p-type bulk
semiconductors\cite{OpticalOrientationBook} and in modulation
p-doped quantum wells.\cite{Damen:1991,Wagner:1993}

\begin{figure}[t]\vspace{0pt}
    \includegraphics[width=7.0cm]{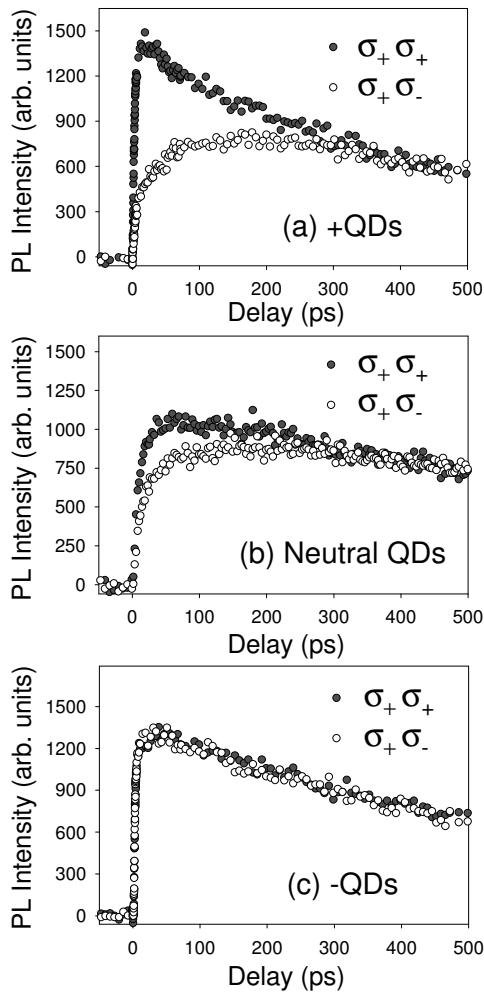}
    \caption{Results of polarization-sensitive, time-resolved photoluminescence experiments on (a) +QDs; (b) neutral QDs;
    and (c) -QDs, showing emission from the ground state optical transition. Filled (open) circles indicated dynamics of the optically injected (minority) carrier spin population following
    capture and relaxation into the QDs.}
    \label{fig:2}
\end{figure}

The self-assembled InAs QDs were grown by molecular beam epitaxy
under identical conditions for structures with and without
modulation doping.  Each sample contains a single layer of QDs
embedded in the center of a 30 nm layer of GaAs, cladded on both
sides with 340 nm AlGaAs barriers.  For all samples, 15 nm of
undoped GaAs was deposited, followed by 2.7 monolayers of InAs to
form the QD layer.  Atomic force microscopy and cross-sectional
transmission electron microscopy on similar structures indicate
that the QDs are 25 nm in diameter and 3 nm in height, with an
areal density of 3$\times$10$^{10}$ cm$^{-2}$. For the neutral QD
structure, 15 nm of undoped GaAs was deposited over the QDs, while
for the modulation doped structures, 12 nm of undoped GaAs was
followed by 3 nm of GaAs doped with Si (-QDs) or Be (+QDs) at a
density of 2$\times$10$^{18}$ cm$^{-3}$.  The associated free
charge carriers accumulate in the lower energy states within the
InAs QDs. Since this doping density corresponds to $\sim$ 20 free
carriers per QD, the QDs are highly charged, in contrast to
previous measurements of spin dynamics in -QDs with a single extra
electron.\cite{Cortez:2002,Kozin:2002} Evidence of substantial
charge accumulation in the QD states is provided by results of
continuous-wave photoluminescence experiments, in which the ground
state interband optical transition is observed to shift by a few
10's of meV to lower (higher) energies for -QDs (+QDs) relative to
the ground state in neutral QDs.\cite{Schmidt:1998,Regelman:2001}

Room temperature time-resolved photoluminescence experiments were
performed with 100 fs, linearly-polarized or circularly-polarized,
1.42 eV pulses from a Ti:sapphire laser tuned to excite carriers
near the GaAs band edge.  The optical selection rules for bulk
III-V semiconductors\cite{OpticalOrientationBook} dictate that
circularly-polarized excitation at the GaAs band edge will produce
electrons and holes with a net spin polarization of 50\%. The
photoluminescence from the ground state optical transition in the
QDs is passed through a quarter waveplate and time-resolved using
sum frequency generation in a KNbO$_{3}$ crystal with a
linearly-polarized Ti:sapphire pulse. The type I phase matching
condition, which permits detection of only one linear polarization
component of the photoluminescence and therefore acts as a linear
polarizer, permits time-resolution of either the emission from
carriers of the same spin as those injected or the opposite spin.
For our experiments, the excited carrier density corresponds to
$\sim$ 1 electron-hole pair per QD, as estimated using the
measured fluence and the absorption coefficient of bulk GaAs at
1.42 eV.\cite{GaAsabscoeff}

Results of polarization-sensitive, time-resolved photoluminescence
experiments on charged and neutral QD structures are shown in
Fig.~\ref{fig:2}(a)-Fig.~\ref{fig:2}(c).  Due to the
spin-sensitive optical selection rules in self assembled InAs QDs
for light emission along the (001) growth
direction,\cite{Pryor:2003} the degree of circular polarization of
the QD photoluminescence, given by:
\begin{equation}
\rho (\%) = 100 \times \frac{\sigma_{+} - \sigma_{-}}{\sigma_{+} +
\sigma_{-}},
\end{equation}
directly reflects the spin polarization of carriers in the QDs.
The results in Fig.~\ref{fig:2} illustrate the strong effect of
the QD charge on the size of the carrier spin polarization: the
degree of circular polarization is 4$\times$ larger in +QDs than
in neutral QDs, while for -QDs, no circular polarization is
observed for any time delay.

\begin{figure}[t]\vspace{0pt}
    \includegraphics[width=7.0cm]{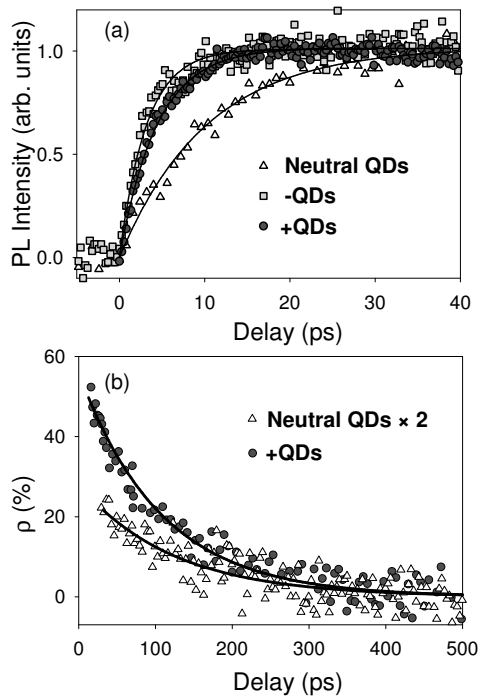}
    \caption{(a) Results of time-resolved photoluminescence experiments under excitation with
linearly polarized, 1.42 eV pulses in order to probe the
spin-independent carrier capture rates in charged and neutral QDs.
Circles: +QDs; triangles: neutral QDs; squares: -QDs.  In neutral
QDs, the photoluminescence reaches a maximum at 45 ps, while in
charged QDs, this maximum occurs earlier (13 ps, -QDs; 18 ps,
+QDs); (b) Decay of degree of circular polarization in +QDs and
neutral QDs. The data for neutral QDs is multiplied by a factor of
two for clarity. Single exponential fits to these data (solid
curves) indicate that the circularly-polarized emission decays on
similar time scales in +QDs ((120 $\pm$ 8) ps) and neutral QDs
((110 $\pm$ 5) ps).}
    \label{fig:3}
\end{figure}

The contrasting amplitudes of the circularly-polarized emission
for the charged and neutral QDs is caused by the influence of the
built-in carrier population on: \textit{(i)} the species of
carrier captured (electron or hole); and \textit{(ii)} the carrier
capture rate.  Due to the large density of carriers available from
the doped GaAs layer adjacent to the QDs, the QD ground state will
be full of electrons (-QDs) or holes (+QDs) prior to optical
excitation in the GaAs. Photoluminescence emission from the ground
state optical transition of charged QDs therefore originates from
capture and relaxation of only the opposite charge carriers.  The
degree of circular polarization of the photoluminescence from +QDs
(-QDs) thus provides an indication of the residual spin
polarization of captured electrons (holes). In contrast for
neutral QDs, emission results from the capture and relaxation of
both electrons and holes.  The examination of carrier spin
dynamics in modulation doped QDs with photoluminescence techniques
therefore allows for the unambiguous separation of electron and
hole spin dynamics, unlike resonant pump probe
experiments.\cite{Paillard:2001} The early time dynamics of the
photoluminescence emission from the charged and neutral QDs under
excitation with linearly-polarized pulses (Fig.~\ref{fig:3}(a))
indicates that carriers are captured more rapidly into charged QDs
than neutral QDs.  This result is attributed to a carrier capture
process mediated by interaction with the built-in carriers in
charged
QDs.\cite{CommentNextQDPaper,Sosnowski:1998,noteAboutExcitedState}
The spin-independent carrier dynamics in these modulation-doped
QDs will be described elsewhere.\cite{CommentNextQDPaper}

As shown in Fig.~\ref{fig:3}(b), the rapid electron capture into
+QDs leads to a substantially larger spin signature than in
neutral QDs: At the time delay corresponding to the peak of the QD
photoluminescence, the degree of circular polarization is 50\% in
+QDs, compared to 10\% in neutral QDs. The observation of 50\% in
+QDs, which is equal to the spin polarization of the carrier
distribution initially injected into the bulk GaAs surrounding the
QDs, indicates that no electron spin information is lost in the
GaAs during the time required for capture into +QDs.  In contrast,
more than three quarters of the initial electron spin polarization
has been lost during the longer 45 ps photoluminescence rise time
in neutral QDs. The absence of circularly-polarized
photoluminescence from the -QDs indicates that hole spins are
randomized in the bulk GaAs prior to capture.
\cite{Kalevich:2001,OpticalOrientationBook} This result also
indicates that the circularly-polarized emission observed in
neutral QDs may be attributed exclusively to the residual spin
polarization of captured electrons.

Single exponential fits to the decay of the circularly-polarized
emission, which appear as the solid curves in Fig.~\ref{fig:3}(b),
provide decay times of
 (120 $\pm$ 8) ps
and (110 $\pm$ 5) ps for the neutral QDs and +QDs, respectively.
Due to the random capture process for carriers into the
QDs,\cite{Grundmann:1997,Zhang:2001} which implies that the
probability for multiple carriers to be captured into a single QD
is significant even for optical excitation of less than a single
electron-hole pair per QD, the decay time of the ground state
circularly-polarized photoluminescence only provides a lower bound
on the spin relaxation time of electrons in the QDs.  This decay
time reflects the combined dynamics of spin relaxation and spin
redistribution among the QD ground and excited
states.\cite{Paillard:2001,Kalevich:2001} The decay time of $\rho$
was observed to increase sharply with decreasing
optically-injected carrier density down to the lowest densities
accessible in these experiments (1 electron-hole pair per QD),
indicating that the effect of capture of opposite spin electrons
is significant. Nevertheless, the decay times we observe are in
line with previous room temperature measurements on neutral
QDs.\cite{Paillard:2001} The most remarkable aspect of the results
of Fig.~\ref{fig:3}(b) is the observation of similar decay times
of the circularly-polarized photoluminescence in neutral QDs and
+QDs. It has been suggested that the presence of holes leads to an
efficient spin relaxation channel for electrons in QDs through the
electron-hole exchange interaction.\cite{Imamoglu:1999} Indeed,
excess holes were found to strongly reduce the electron spin decay
time in p-doped bulk semiconductors\cite{OpticalOrientationBook}
and in p modulation-doped quantum
wells.\cite{Damen:1991,Wagner:1993}  The room temperature
experimental results we present here, which represent the first
measurement of electron spin dynamics in positively-charged QD
nanostructures, indicate that the presence of even a large
population of excess holes has little if any influence on the
electron spin dynamics.

In order to examine the implications of the higher electron spin
polarization in +QDs for a spin detection application such as a
spin LED,\cite{Fiederling:1999,Ohno:1999,Chye:2002} we evaluated
the steady state circular-polarized luminescence for +QDs and
neutral QDs by numerically integrating the measured ground state
photoluminescence over time delay for the two polarization
geometries.\cite{note:integration} These findings indicate that
the rapid spin-polarized electron capture into +QDs leads to a
sixfold increase in the steady-state circularly-polarized
emission, demonstrating a substantial improvement in the electron
spin detection efficiency in QDs with the incorporation of a large
density of built-in holes.

In summary, we have applied time- and polarization-resolved
photoluminescence up-conversion to the study of spin-polarized
carrier dynamics in charged and neutral InAs QDs at room
temperature. Compared to neutral QDs, shorter carrier capture and
relaxation times were measured in charged QDs, attributed to
electron-hole interactions involving the built-in carriers. In
+QDs, these electron-hole scattering processes lead to rapid
transfer of optically-injected spin-polarized electrons from the
bulk GaAs to the QD ground state, resulting in a six-fold
enhancement in the time-integrated circularly-polarized
photoluminescence compared to neutral QDs. Measurements of spin
decay dynamics in +QDs and neutral QDs indicate that the large
built-in hole population in the +QDs has little effect on the
electron spin relaxation time. These factors, together with the
high radiative efficiency of QDs, indicates that LEDs
incorporating positively charged QDs offer superior performance as
opto-electronic detectors of spin-polarized electrons.

This research is supported by the DARPA MDA972-01-C-0002 and the
Natural Sciences and Engineering Research Council of Canada.


\begin{references}
\bibitem{noteAboutExcitedState} The measured rise
time of the photoluminescence for the excited state interband
transition, which occurs 60-85 meV above the ground state
interband transition in these QDs, is similar to the ground state
results of Fig.~\ref{fig:3}(a), allowing the influence of the
built-in carrier population on the carrier capture dynamics to be
separated from the the dynamics of inter-level relaxation.
\bibitem{note:integration} In performing this
integration, we utilized the analytic exponential fit to the
long-term photoluminescence decay, corresponding to the
recombination time of the photogenerated carriers, in order to
extend the integration interval beyond the 3 ns measured delay
range.
\end{references}
\end{document}